\documentclass[12pt]{article}
\begin{document}

\title{\bf The Pseudo-Newtonian Force and Potential
of the Stringy Black Holes}

\author{M. Sharif \thanks{e-mail: msharif@math.pu.edu.pk}
\\ Department of Mathematics, University of the Punjab,\\
Quaid-e-Azam Campus Lahore-54590, PAKISTAN.}

\date{}

\maketitle
\begin{abstract}
This paper is devoted to investigate the structure of the
pseudo-Newtonian force and potential of the stringy black holes. We
discuss conditions for the force character from an attractive to
repulsive. It is also found that the force will reach a maximum
under certain conditions. Also, the ratio of mass and charge is
evaluated for the maximum force.
\end{abstract}

General Relativity replaces the bending of paths by the curvature of
spacetime rather than as caused by a force. While retaining the path
predicted by general relativity, we can flatten the background
spacetime and ask what force gives the required curvature. This
relativistic analogue of the gravitational force has been called the
pseudo-Newtonian ($\psi N$) force. For a point gravitating source,
it just gives the usual Newtonian gravitational force. In the $\psi
N$ approach [1-3], the curvature of the spacetime is {\it
straightened out} to yield a relativistic force which bends the
path, so as to again supply the guidance of the earlier,
force-based, intuition.

The quantity whose gradient gives the $\psi N$ force is the $\psi N$
potential. Since the $\psi N$ force depends on the choice of
geodesic the procedure becomes unwieldy for more complicated cases.
However, there is a well-defined class of observers for whom the
$\psi N$ force can be shown to be the gradient of a scalar $\psi N$
potential [4],
\begin{equation}
V=\frac{1}{2}(\textbf{k}.\textbf{k}-1),
\end{equation}
where $\textbf{k}$ is the Killing vector corresponding to the
timelike isometry. This $V$ is exactly the usual conjecture for the
gravitational potential [5]. The $\psi N$ force is the
generalisation of the force which gives the usual Newtonian force
for the Schwarzschild metric and a $"\frac{Q^2}{r^3}"$ correction to
it in the Riessner-Nordstrom metric. The $\psi N$ force may be
regarded as the {\it Newtonian fiction} which {\it explains} the
same motion (geodesic) as the {\it Einsteinian reality} of the
curved spacetime does. We can, thus, translate back to Newtonian
terms and concepts where our intuition may be able to lead us to
ask, and answer, questions that may not have occurred to us in
relativistic terms. The structure of the $\psi N$ force and
potential for a highly charged and rotating black hole is very
interesting. Infact the $\psi N$ force itself reaches a maximum and
then starts to decrease outside the black hole. Thus an observer
armed with an accelerometer could deduce the presence of the black
hole and its nature while still able to communicate with the
outside.

Some insights have already been obtained [1,4] by expressing the
consequences of general relativity in terms of forces by applying it
to Kerr and Kerr-Newmann metrics. Ivanov and Prodanov [6] have
studied the pseudo-Newtonian potential for charged particle in
Kerr-Newman geometry. In recent papers, we have investigated the
structure of the pseudo-Newtonian force and potential about a five
[7] and $n$-dimensional [8] rotating black holes. Here we analyse
the structure of force and potential of the stringy black holes.

In the free fall rest-frame, the $\psi N$ force is given [2,3] by
\begin{equation}
F_i=-M(\ln \sqrt {g_{00}})_{,i},\quad (i=1,2,3).
\end{equation}
This can be written as
\begin{equation}
F_i=-V_{,i},
\end{equation}
where
\begin{equation}
V=M(\ln \sqrt {g_{00}}).
\end{equation}
It is clear that $V$ is the generalisation of the classical
gravitational potential and, for small variations from Minkowski
space
\begin{equation}
V\approx\frac{1}{2}M(g_{00}-1)
\end{equation}
which is the pseudo-Newtonian potential. We shall investigate the
behaviour of these quantities for the stringy black holes.

The non-trivial geometry of different types of black holes is of
crucial interest in general relativity. We shall consider some
typical four dimensional stringy black hole spacetimes. In four
dimensions, the simplest eternal black hole geometry is, of course,
the Schwarzschild. We consider the variations of the Schwarzschild
which have arisen in the context of string theory [9-11]. All these
four dimensional geometries have a Schwarzschild limit (obtainable
by setting a parameter in the line element to zero). We also have
asymptotically flat solutions representing black holes in
dilaton-Maxwell gravity. Such solutions, due to Garfinkle, Horowitz
and Strominger [9], represent electric and dual magnetic black holes
[10]. The spacetime geometry of these line elements are causally
similar to the Schwarzschild geometry. In the respective limit of
$\alpha=0$ or $Q=0$, we have Schwarzschild geometry in both the
cases.

The metric for the black hole with electric charge is given by
\begin{eqnarray}
ds^2=\frac{(1-\frac{2m}{r})}{(1+\frac{2m\sinh^2\alpha}{r})^2}dt^2
+\frac{dr^2}{(1-\frac{2m}{r})}+r^2d\Omega^2,
\end{eqnarray}
where $d\Omega^2=d\theta^2+\sin^2\theta d\phi^2$ is the metric on a
2-dimensional unit sphere and $\alpha$ is a parameter related to the
electric charge of the black hole such that
$\tanh^2\alpha=\frac{Q^2}{M^2},~m$ is a parameter related to the
physical mass of the black hole. For $\alpha=0$, the spacetime
reduces to the Schwarzschild spacetime. Also, it reduces to
Minkowski spacetime for $m=0$.

The structure of the $\psi N$ force (per unit mass of the test
particle) for the black hole with electric charge takes the
following form
\begin{equation}
F_r=-\frac{\frac{m}{r^2}(1+\sinh^2\alpha-\frac{2m}{r}\sinh^2\alpha)}
{(1-\frac{2m}{r})(1+\frac{2m}{r}\sinh^2\alpha)},\quad F_\theta=0.
\end{equation}
If we take $\alpha=0$, the results exactly coincide with those
$a=0=Q$ in [4], i.e. for the Schwarzschild black hole. Equation (7)
shows that the radial component cannot be zero outside the horizon.
Consequently, the attractive force cannot change to a repulsive
force outside the black hole. It is obvious that naked singularities
can give repulsive as well as attractive forces [5]. The force
structure would provide interesting features if it reaches a maximum
and then drops as we reduce $r$ provided the turnover lies outside
the horizon. Since our observers are seeing {\it force} in a flat
space, the metric to be used is the plane polar one. Thus the
magnitude of the force is
\begin{equation}
F=\frac{\frac{m}{r^2}(1+\sinh^2\alpha-\frac{2m}{r}\sinh^2\alpha)}
{(1-\frac{2m}{r})(1+\frac{2m}{r}\sinh^2\alpha)}.
\end{equation}
The equation for the turnover along radial direction is
\begin{eqnarray*}
(1+2\sinh^2\alpha)r^3-m(1+4\sinh^2\alpha+2\sinh^4\alpha)r^2\nonumber\\
+4m^2\sinh^2\alpha(1-\sinh^2\alpha)r+4m^3\sinh^3\alpha=0.
\end{eqnarray*}
This is a cubic equation. In general a cubic equation can be solved
and we can have the following three different possibilities:\\
(i) All the three roots are real and distinct.\\
(ii) All the three roots are real, two of them being equal.\\
(iii) One root is real and the other two are a conjugate pair. When
we solve the above cubic equation, we obtain one real and two
imaginary roots. The real root is given by
\begin{eqnarray}
r&=&\frac{1}{3\cosh{2\alpha}}[\{m+m(3+\cosh{2\alpha})\sinh^2{\alpha}
+(m^2(1-4\sinh^2{\alpha}\nonumber\\
&+&8\sinh^4{\alpha}+40\sinh^6{\alpha}+4\sinh^8{\alpha}))\}/
[\{6\sqrt{3}\{(-m^6\cosh^2{2\alpha}\sinh^3{\alpha}\nonumber\\
&\times&(1+\sinh{\alpha}(1+\sinh{\alpha}(-6+\sinh{\alpha}(-37
+\sinh{\alpha}(-36+\sinh{\alpha}(-87\nonumber\\
&+&4\sinh{\alpha}(10+\sinh{\alpha} (-19+\sinh{\alpha}
(54+\sinh{\alpha}(-22+\sinh{\alpha}(30\nonumber\\
&+&\sinh{\alpha}(10+\sinh{\alpha}
(2+\sinh{\alpha}))))))))))))))\}^\frac{1}{2}+m^3(1+2\sinh^2{\alpha}
(-3\nonumber\\
&+&\frac{1}{64}\sinh{\alpha}
(-864-864\cosh{4\alpha}+234\sinh{\alpha}-664\sinh{3\alpha}
+48\sinh{5\alpha}\nonumber\\
&+&51\sinh{7\alpha}+\sinh{9\alpha})))\}^\frac{1}{3}]+\{6\sqrt{3}
\{(-m^6\cosh^2{2\alpha}\sinh^3{\alpha}(1+\sinh{\alpha}\nonumber\\
&\times&(1+\sinh{\alpha}(-6+\sinh{\alpha}(-37+\sinh{\alpha}
(-36+\sinh{\alpha}(-87+4\sinh{\alpha}(10\nonumber\\
&+&\sinh{\alpha}(-19+\sinh{\alpha}(54+\sinh{\alpha}(-22+\sinh{\alpha}
(30+\sinh{\alpha}(10\nonumber\\
&+&\sinh{\alpha}
(2+\sinh{\alpha}))))))))))))))\}^\frac{1}{2}+m^3(1+2\sinh^2{\alpha}
(-3\nonumber\\
&+&\frac{1}{64}\sinh{\alpha}
(-864-864\cosh{4\alpha}+234\sinh{\alpha}\nonumber\\
&-&664\sinh{3\alpha}+48\sinh{5\alpha}+51\sinh{7\alpha}+\sinh{9\alpha})))\}^\frac{1}{3}].
\end{eqnarray}
This equation is not easy to analyse analytically. The general
comments can be given as follows. We see that a maximum of the
magnitude does occur at this value of $r$. The maximum value of the
force can be obtained by replacing this value of $r$ in Eq.(8) which
turns out to be a complicated equation. However, this will obviously
yield that the maximum value of the force depends on the value of
$\alpha$ which corresponds to charge. This will provide turnovers
inside the horizon and also on the surface of the horizon for an
extreme black hole.

It would be interesting to explore the approximate solutions for
sufficiently small charge. For this purpose, we expand $\sinh\alpha$
and $\cosh\alpha$ up to first order and neglecting the second and
higher orders. Using this approximation, Eq.(8) gives
\begin{equation}
F=\frac{\frac{m}{r^2}(1+\alpha)}{(1-\frac{2m}{r})}.
\end{equation}
The maximum value turns out to be at $r=m$ and the corresponding
value will be
\begin{equation}
F^*=-\frac{(1+\alpha)}{m},\quad m>8/3,\quad 1+\alpha<0.
\end{equation}
This would yield the condition on the ratio of mass and charge to
have the maximum force for this black hole. It is worthwhile to
mention here that these limits on the ratio appear in the discussion
of the pseudo-Newtonian force of the charged Kerr metric [4] as well
of critical accretion [12].

The general results for particular choice of charge and mass (using
Eqs.(8) and (9)) can be summarized in the following table.\\\\
\textbf{Table 1.} Maximum Force for particular values of Charge and
Mass
\begin{center}
\begin{tabular}{|c|c|c|c|c|c|c|c|c|c|}
\hline   \textbf{$m$} & \textbf{$Q$} & \textbf{$r$} & \textbf{$F^*$}\\
\hline 2 & 0.1 & 1.98898 & 0.496247 \\
\hline 2 & 0.02 & 1.95128 & 0.484949\\
\hline 2.5 & 0.1 & 0.49135 & 1.09407\\
\hline 2.5 & 0.2 & 2.4626 & 0.392304\\
\hline 2.5 & 0.15 & 2.47978 & 0.395675\\
\hline 3 & 0.025 & 2.999958 & 0.333264\\
\hline 5 & 0.025 & 4.99975 & 0.199985\\
\hline
\end{tabular}
\end{center}
This provides inverse behavior for maximum force corresponding to
the stationary point.

The corresponding potential is found using Eqs.(4) and (6) given by
\begin{equation}
V=-\frac{\frac{m}{r}(1+\frac{2m\sinh^4\alpha}{r}+2\sinh^2\alpha)}{2(1+\frac{2m\sinh^2\alpha}{r})}.
\end{equation}
We note that the structure of force and potential indicate similar
type of behaviour as for the charged Kerr metric [4].

The dual (magnetic) metric of Eq.(6) is given as follows
\begin{eqnarray}
ds^2&=&\frac{(1-\frac{2m}{r})}{(1-\frac{Q^2}{mr})}dt^2
+\frac{dr^2}{(1-\frac{2m}{r})(1-\frac{Q^2}{mr})}+r^2d\Omega^2,
\end{eqnarray}
where $Q$ is the magnetic charge of the black hole.

The structure of the $\psi N$ force (per unit mass of the test
particle) for this black hole takes the following form
\begin{equation}
F_r=-\frac{\frac{m}{r^2}(1-\frac{Q^2}{2m^2})}{2r(1-\frac{2m}{r})(1-\frac{Q^2}{mr})},\quad
F_\theta=0.
\end{equation}
We see that the radial component can never become zero outside the
horizon and so the force cannot change character from an attractive
to a repulsive force outside the black hole. The magnitude of the
force is
\begin{equation}
F=\frac{\frac{m}{r^2}(1-\frac{Q^2}{2m^2})}{2r(1-\frac{2m}{r})(1-\frac{Q^2}{mr})}.
\end{equation}
The equation for the turnover along radial direction is
\begin{equation}
2mr-Q^2-2m^2=0
\end{equation}
satisfied for the value of $r$ given by
\begin{equation}
r=\frac{Q^2+2m^2}{2m}.
\end{equation}
We see that a maximum of the magnitude does occur at this value of
$r$. Thus the maximum value of the force is
\begin{equation}
F^*=\frac{m}{2m^2-Q^2}
\end{equation}
which indicates the effect of $Q$ on the force. It is easy to find
the ratio of mass to charge from here. The corresponding potential
is obtained by using Eqs.(4) and (12)
\begin{equation}
V=-\frac{\frac{m}{r}(1-\frac{Q^2}{2m^2})}{(1-\frac{Q^2}{mr})}.
\end{equation}
Notice that the structure of force and potential coincide with that
of the Kerr metric [4]. The general expression for the stationary
point of the first metric does not provide much insight. However, we
have explored it for small value of $\alpha$ for the purpose of
discussion. We would like to point out that the charges and
rotations can produce repulsive force. This has been verified for
charged spacetimes using this formulation. It is interesting to
mention here that for $\alpha=0=Q$, the results of the force and
potential for both the spacetimes reduce to the Schwarzschild
metric.

\vspace{1cm}

{\bf \large References}

\begin{description}

\item{[1]} Qadir, A. and Quamar, J.: {\it Proc. 3rd Marcel Grossmann Meeting on
General Relativity}, ed. Hu Ning (North Holland Amstderm 1983)189;\\
Quamar J.: {\it Ph.D. Thesis} Quaid-i-Azam University Islamabad
(1984).

\item{[2]} Qadir, Asghar and Sharif, M.: Nuovo Cimento B {\bf 107}(1992)1071;\\
Sharif, M.: {\it Ph.D. Thesis} Quaid-i-Azam University Islamabad
(1991).

\item{[3]} Qadir, Asghar and Sharif, M.: Phys. Lett. A {\bf 167}(1992)331;\\
Sharif, M.: Astrophys. and Space Science {\bf 253}(1997)195.

\item{[4]} Qadir, A. and Quamar, J.: Europhys. Lett. {\bf
2}(1986)423;\\
Qadir, A.: Europhys. Lett. {\bf 2}(1986)427.

\item{[5]} Hawking, S.W. and Ellis, G.F.R.: \emph{The Large Scale Structure of
Spacetime} (Cambridge University Press, 1973).

\item{[6]} Ivanov, Rossen I. and Prodanov, M.: Phys. Lett. B {\bf611}(2005)34.

\item{[7]} Sharif, M.: Nuovo Cimento B {\bf121}(2006)121.

\item{[8]} Sharif, M.: Nuovo Cimento B {\bf122}(2007)343.

\item{[9]} Garfinkle, D., Horowitz, G.T. and Strominger, A.:
Phys. Rev. D \textbf{43}(1991)3140; Erratum \emph{ibid.} Phys. Rev.
D \textbf{45}(1992)3888.

\item{[10]} Horowitz, G.T.: The Dark Side of String Theory: Black Holes
and Black Strings, hep-th/9210119.

\item{[11]} Green, M.S., Schwarz, J.H. and Witten, E.: \emph{Superstring Theory}
(Cambridge University Press, 1987);\\
Polchinski, J.: \emph{String Theory} (Cambridge University Press,
1997).

\item{[12]} Jamil, M., Qadir, A. and Rashid, M.A: Eur. Phys. J. {\bf 58C}(2008)325.
\end{description}

\end{document}